\documentclass[preprint]{revtex4-2}
\usepackage{graphicx, color}  
\usepackage{dcolumn}   
\usepackage{bm}        
\usepackage{amssymb}   
\usepackage{amsmath}   
\usepackage{footmisc}
\usepackage{enumerate}
\usepackage{comment}
\usepackage{tabularx}
\usepackage{booktabs}
\usepackage{placeins}
\usepackage{cancel}

\usepackage{hyperref}
\usepackage{siunitx}

\newcommand{\RN}[1]{%
  \textup{\uppercase\expandafter{\romannumeral#1}}%
}
\begin{document}


\title{$B_c$, $B_s$ and $D_s$ production at lepton-hadron colliders}

\author{Qi-Wei Hu$^{1}$}
\author{Cong-Feng Qiao$^{1}$}
\email[Corresponding author:]{qiaocf@ucas.ac.cn}
\author{Li-Ping Sun$^2$}

\address{$^{1}$School of Physics, University of Chinese Academy of Sciences, Beijing 100049, China}
\address{$^2$School of Science, Beijing University of Civil
 Engineering and Architecture, Beijing 100044, China}
\author{~\\~\\~\\}

\begin{abstract}
\vspace{0.5cm}
Within the framework of nonrelativistic quantum chromodynamics, this study examines the electroproduction processes $e+p\to e+B_c+\overline{c}+b$, $e+p\to e+B_s+\overline{s}+b$, and $e+p\to e+D_s+\overline{c}+s$ at lepton-hadron colliders. The differential cross sections in $\cos\theta$ and $p_T^2$ at HERA are presented. The results indicate that the production of $B_c$ is feasible at the EIC, whereas it is not at HERA. The cross sections for $B_s$ and $D_s$ are notably significant at HERA, yet they exhibit sensitivity to variations in quark mass and renormalization scale.

\end{abstract}

\maketitle
\newpage
\section{Introduction}

The field of heavy quarkonium research originated with the 1974 discovery of charmonium $J/\psi$ \cite{E598:1974sol,SLAC-SP-017:1974ind}, followed by the 1977 identification of bottomium $\Upsilon$ at Fermilab\,\cite{E288:1977xhf}. Heavy quarkonium serves as an exemplary system for quantum chromodynamics (QCD) research. The low relative velocity of quarks within heavy quarkonium allows for the factorization of scattering amplitudes into hadronization and perturbative components. Based on this property, nonrelativistic quantum chromodynamics (NRQCD) \cite{Bodwin:1994jh} is pivotal in addressing the production and decay processes of these systems. 

The $B_c$ meson stands out among mesons for its composition of two distinct heavy flavor quarks, leading to decay processes governed solely by weak interactions. This characteristic offers a unique opportunity to probe quark weak interactions. Despite its small production cross section, the S-wave ground state $B_c({ }^1 S_0)$ was first observed in 1998 at Tevatron by the CDF collaboration \cite{CDF:1998ihx}. Within the $B_c$ family, the S-wave vector state $B_c^*({ }^3 S_1)$\,\cite{ATLAS:2014lga} decays predominantly into $B_c$, underscoring the significance of $B_c^*$ in $B_c$-related processes.


Production of $B_c$ can be categorized as either direct or indirect. Direct production refers to production at colliders by partons, $e^\pm$, $\gamma^{(*)}$, etc. Indirect production involves the decay of massive intermediate particles such as $W^\pm$, $Z$, $H$, and $t$ into $B_c$. At $ep$ colliders, $B_c$ production is further differentiated into photoproduction and electroproduction. Photoproduction involves the initial photon being emitted through electron bremsstrahlung, with energy distribution described by the Weizsäcker-Williams approximation \cite{Frixione:1993yw}. Laser back scattering (LBS) can generate photons with higher energy due to significant electron energy transfer \cite{Ginzburg:1983qc}. Electroproduction, which involves electron interaction with a gluon via a virtual photon, is the focus of this study.

The LHC, with its record collision energy and extensive data, has made strong production a focal point since its inception. The leading order (LO) calculations for $B_c$ strong production, which began in the 1990s \cite{Chang:1992jb,Chang:1994aw}, have concentrated on S-wave $B_c$ production at hadron colliders, including the Tevatron and LHC. For higher excited state $B_c$ mesons under the NRQCD framework, Ref. \cite{Chang:2004bh} calculated P-wave $B_c$ strong production, while Ref. \cite{Chang:2005bf} examined the color octet mechanism (COM) in $B_c$ strong production. Ref. \cite{Chang:2005wd} also provided a detailed analysis of $B_c$ strong production driven by intrinsic heavy quarks within colliding hadrons. Note, the heavy quark fragmentation method \cite{Mele:1990cw,Braaten:1993jn} is also widely used in production of mesons with $b$ or $c$ quark, including $B$, $D$ and $D_s$ families \cite{Kniehl:2005ej,Cacciari:2012ny}. 

Beyond strong production, $B_c$ can also arise from electron-positron collisions, electron-proton collisions, and heavy-ion collisions. Ref. \cite{Yang:2011ps} analyzed $B_c$ production in $Z$ factory, encompassing S-wave, P-wave, and color octet states. Refs. \cite{Kolodziej:1994uu,Berezhnoy:1995ay} calculated the double photon production of S-wave and P-wave $B_c$, revealing that differential distributions under NRQCD and fragmentation methods are markedly different. At $ep$ colliders, $B_c$ can be produced through interactions involving photons and gluons \cite{Berezhnoy:1997er}. At the LHC, heavy-ion Pb + Pb collisions can also yield $B_c$ \cite{Liu:2012tn,Norbeck:2013zba}.

Compared to $pp$ colliders, $ep$ colliders operate at lower energies but offer higher luminosities and distinct scientific goals. The HERA \cite{H1:2010udv} has ceased operations, while the EIC \cite{AbdulKhalek:2021gbh,Accardi:2012qut} and EIcC \cite{Anderle:2021wcy} are currently under construction. The EIC, in contrast to HERA, operates at lower energies but with higher luminosity and is capable of producing highly polarized electron and proton beams. Its design objectives include using electrons as probes to investigate the internal structure of nucleons. The results presented in this article are based on these $ep$ colliders.

This paper aims to fill a current gap in the electroproduction of $B_c$ and provide an estimate valuable for future collider experiments. Additionally, the production of $B_s$ and $D_s$ mesons are calculated to evaluate the applicability of NRQCD framework in mesons containing a single heavy quark.

\section{Formalism}\label{sec:experiment}

In NRQCD, the cross section can be decomposed as follows:
\begin{eqnarray}
\label{eqn:nrqcd}
   \mathrm{d}\sigma (ep\rightarrow e+ M+ q+ \bar{q}^\prime)=\int \mathrm{d}xf_{g/p} (x,\,\mu^2)\nonumber\\
   \times \mathrm{d}\sigma (eg\rightarrow e+q\bar{q}^\prime\left [ n \right ]+ q+ \bar{q}^\prime) \langle{\mathcal{O}}^M(n)\rangle\,,
\end{eqnarray}
where $M$ is the final meson, $q$ and $\bar{q}^\prime$ is the component quark of meson, $\langle \mathcal{O}^M(n) \rangle$ represents the long-distance matrix element (LDME), $n={}^{2S+1}L_J$ is the standard spectroscopic notation for the quantum number of the $q\bar{q}^\prime$ pair, and $f_{g/p} (x,\,\mu^2)$ is the parton distribution function (PDF) of gluon evaluated at the factorization scale $\mu^2$.


Furthermore, for electroproduction, the leptonic tensor formula \cite{Sun:2017nly} can formally reduce the number of Mandelstam invariants:
\begin{flalign}
\label{eqn:lep}
    &\mathrm{d}\sigma (eg\rightarrow e+q\bar{q}^\prime[n]+ q+ \bar{q}^\prime)=\nonumber\\
    &\frac{1}{4xP\cdot k}\frac{1}{N_cN_s}\frac{1}{(Q^2)^2}L_{\mu \nu }H^{\mu \nu }[n]\mathrm{dPS}_4\,.
\end{flalign}
The notation follows the reference, $N_c$ and $N_s$ are the color and spin number, $P$ and $k$ are momentum of initial proton and electron, $Q^2$ is the virtuality of $\gamma^*$, $L_{\mu \nu}$ is the leptonic tensor for the initial electron emitting a virtual photon, $H^{\mu \nu}[n]$ is the hadronic part of the interaction between the virtual photon and the gluon, and $\mathrm{dPS}_4$ is the $4$ body phase space for $e+B_c+ b+ \bar{c}$. To improve the efficiency, the phase space can be decomposed:
\begin{flalign}
\label{eqn:phase}
    &\int \mathrm{dPS}_4 =
    (2\pi)^2\int_{4m^2}^{E^2}\mathrm{d}P_{B_cb\bar{c}}^2 \mathrm{dPS}_2(e + g \rightarrow e + [B_cb\bar{c}])\times \nonumber\\
    &\int_{m^2}^{M^2}\mathrm{d}P_{b\bar{c}}^2 \mathrm{dPS}_2([B_cb\bar{c}] \rightarrow B_c + [b\bar{c}])\mathrm{dPS}_2([b\bar{c}] \rightarrow b + \bar{c})\,,
\end{flalign}
where $\mathrm{dPS}_2$ is the two body phase space, $m=m_b+m_c$, E is the collision energy, $P_X$ is the four momentum of the system $X$, and $M=\sqrt{P_{B_cb\bar{c}}^2}-m$.

The $q\bar{q}^\prime$ pair hadronization process can be computed using the covariant projection operator method. For pseudoscalar, the projection operator can be expressed as:
\begin{eqnarray}
\label{eqn:projection}
\frac{1_c}{\sqrt{N_c}} \otimes \frac{1}{2\sqrt{M}} \gamma_5 (\cancel{P} + m_1 + m_2)\,,
\end{eqnarray}
where $1_c$ is the identity matrix in color space, $M$ and $P$ are the mass and momentum of the meson, respectively, and $m_1$ and $m_2$ are the masses of the constituent quarks. For vector mesons, just replace $\gamma_5$ with $\cancel{\epsilon}$.

\begin{figure}[t!]
    \includegraphics[width=0.75\textwidth]{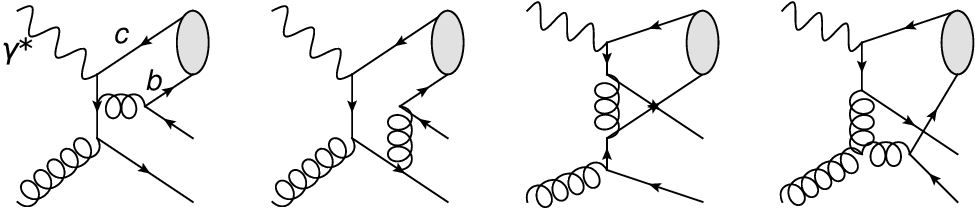}
    \caption{Typical LO Feynman diagrams of $\gamma^*+g\rightarrow B_c+\overline{c}+b$.}
\label{fig:feyndiag}
\end{figure}

Typical LO Feynman diagrams are shown in Fig. \ref{fig:feyndiag}. All the 24 diagrams can be generated by varying the gluon and photon, reversing fermion lines, and considering different quark pairs for the meson. For $bc$ and $cs$ pairs, the same analytical amplitude are used, adjusting only the corresponding quark masses during the Monte Carlo simulation. For $bs$ pair, although the 24 diagrams appear identical to those of $bc$ and $cs$, the color algebra structure differs, resulting in a distinct analytical amplitude.

The following packages are applied in calculations:
\begin{itemize}
    \item FeynArts \cite{Kublbeck:1990xc}: for generating Feynman diagrams
    \item FeynCalc \cite{Mertig:1990an}: for generating analytical amplitude expressions and performing SU(3) color algebra
    \item CUBA \cite{Hahn:2004fe}: for numerically calculating the phase space integral
\end{itemize}
Feyncalc seems to give wrong results when contracting long chains of Dirac gamma matrix, and can be solved by manually add contracting rules.

\section{Results}\label{sec:result}

The electron mass and the mass split between vector and pseudoscalar mesons are neglected in calculation. Quark masses are set as
\begin{eqnarray}
\label{eqn:quarkmass}
    m_c=1.5\,\text{GeV},\,m_b=4.9\,\text{GeV},\, m_s=0.5\,\text{GeV},
\end{eqnarray}
for $B_c$, $B_s$, and $D_s$.

The running coupling constant uses the one-loop formula at LO. For charm mesons, the number of quark flavors $n_f=4$, and the QCD scale ${\varLambda}_{\text{QCD}}=297\,\text{MeV}$. For bottom mesons, $n_f=5$, and ${\varLambda}_{\text{QCD}}=214\,\text{MeV}$ \cite{ParticleDataGroup:2024cfk}. The renormalization scale $\mu$ is set to $\mu=m_T=\sqrt{M^2+p_T^2}$, where $M$, $m_T$, and $p_T$ are the mass, transverse mass, and transverse momentum of the meson, respectively. To estimate the uncertainty, $\mu$ is set within the range of $\frac{1}{2}m_T<\mu<2m_T$, and the quark mass is adjusted within the range of $\pm0.1\,\text{GeV}$ with the meson mass held constant. Notably, the result specifically isolates the uncertainties arising solely from mass variations. The relationship between the LDME and the wave function at the origin of the meson is $\langle \mathcal{O}^M \rangle = 2(2J+1)N_c{|R(0)|}^2/ 4\pi $. Similar to $J/\psi$, the LDME can be extracted from the meson decay width. However, since the quark flavors are different, the width here is from weak decay, proportional to the corresponding decay constant $f^2$ \cite{ParticleDataGroup:2024cfk}:
\begin{eqnarray}
\label{eqn:weakdecay}
    {\varGamma}^{0}(P\rightarrow l\nu) = \frac{G_F^2}{8\pi} f_P^2 m_l^2 M_P {(1-\frac{m_l^2}{M_P^2})}^2 {|V_{q_1q_2} |}^2\,,
\end{eqnarray}
where $M_P$ and $f_P$ is the mass and decay constant of the pseudoscalar meson, $m_l$ is the final lepton mass, $|V_{q_1q_2}|$ is the Cabibbo-Kobayashi-Maskawa (CKM) matrix element between the quarks in the pseudoscalar meson, and $G_F$ is the Fermi coupling constant. The relationship between  $f_P$ and $|R(0)|$ is 
\begin{eqnarray}
\label{eqn:fp}
    f_P^2 = N_c{|R(0)|}^2/\pi M_P\,.
\end{eqnarray}
For $B_c$, the wave function can be directly calculated from the Buchmüller-Tye potential \cite{Eichten:1994gt,Buchmuller:1980su}. For $B_s$ and $D_s$, the wave functions are obtained from the decay constant. PDG\,\cite{ParticleDataGroup:2024cfk} provides the decay constants $f_{Bs} = 0.234\,\text{GeV}$ and $f_{Ds} = 0.245$ GeV, which are determined from a combination of experimental measurements and lattice QCD calculations. Finally, at leading order:
\begin{eqnarray}
\label{eqn:R}
    {| R_{Bc} (0)|}^2 = 1.642\,\text{GeV}^3,\,| R_{B_c^*}(0)| = | R_{Bc} (0)|\,, \nonumber \\
    {| R_{Bs} (0)|}^2 = 0.299\,\text{GeV}^3,\,| R_{B_s^*}(0)| = | R_{Bs} (0)|\,, \nonumber \\
    {| R_{Ds} (0)|}^2 = 0.124\,\text{GeV}^3,\,| R_{D_s^*}(0)| = | R_{Ds} (0)|\,.
\end{eqnarray}
According to NRQCD heavy quark spin symmetry at leading order in $v$ \cite{Bodwin:1994jh}, pseudoscalar and vector mesons share the same $R(0)$. For the PDF, CT10 \cite{Lai:2010vv} is used.

The collision parameters and experimental cuts are taken from HERA: the electron energy is $27.5\,\text{GeV}$, the proton energy is $920\,\text{GeV}$, $p_T > 1\,\text{GeV}$, and $0.3 < z < 0.9$. The electroproduction photon virtuality is set as $2\,\text{GeV}^2 < Q^2 < 100\,\text{GeV}^2$. Given that the production cross section for $B_c$ is much smaller than that for $J/\psi$, $50\,\text{GeV} < W < 250\,\text{GeV}$ is relatively broader than the constraints employed at HERA. $W = \sqrt{(p_\gamma^2 + p_P^2)}$ is the mass of the hadronic final state, $z = (p_1 \cdot p_P) / (p_\gamma \cdot p_P)$ is the elasticity of the meson production process, and $p_1$, $p_P$, $p_\gamma$ are the momenta of the meson, proton, and photon, respectively.

\begin{figure}[t!]
    \includegraphics[width=0.75\textwidth]{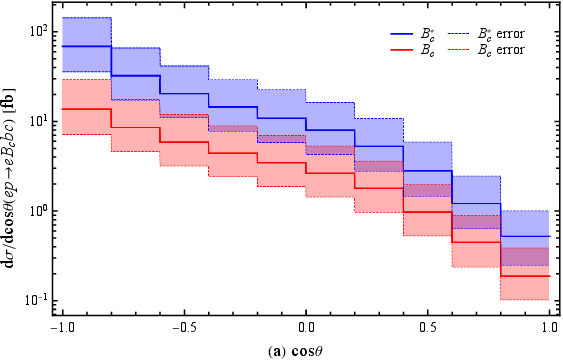}
    \includegraphics[width=0.75\textwidth]{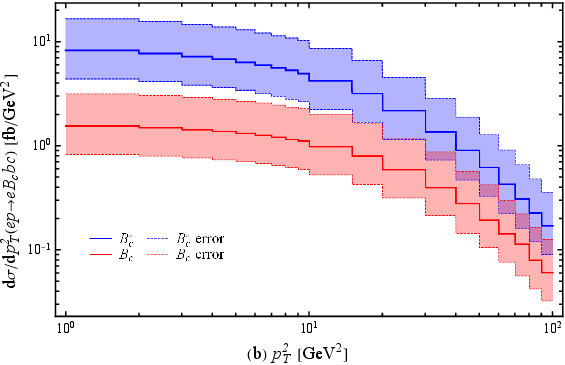}
    \caption{The differential cross section in (a) $\cos\theta$ and (b) $p_T^2$ distributions of the electroproduction process $e+g\to e+B_c+\overline{c}+b$ at LO. Blue lines represent the vector $B_c^{*+}$, and red lines represent the pseudoscalar $B_c^+$. The upper bound corresponds to $m_c=1.4\,\text{GeV}$, $m_b=5.0\,\text{GeV}$, and $\mu=\frac{1}{2}m_T$. The lower bound corresponds to $m_c=1.6\,\text{GeV}$, $m_b=4.8\,\text{GeV}$, and $\mu=2m_T$.}
\label{fig:bc}
\end{figure}

The $\cos\theta$ and $p_T^2$ distributions of the $B_c$ and $B_c^*$ production cross sections are shown in Fig. \ref{fig:bc}, where $\theta$ is the angle between the meson and the incident electron. The cross section of $B_c^*$ is much larger than that of $B_c$, demonstrating its significance in $B_c$ associated processes. As expected, the $\cos\theta$ distribution is concentrated in the incident proton direction $\cos\theta = -1$, and decreases rapidly as $\cos\theta$ increases. To be precise, the area of $\cos\theta \lesssim -0.4$ and $p_T^2 \lesssim 30\,\text{GeV}^2$ is dominant. On the other hand, data at large $\cos\theta$ or $p_T^2$ are too small and may be severely affected by Monte Carlo errors. Nevertheless, disregarding the minor data fluctuations, the $B_c^*$ differential cross section decreases more rapidly compared to that of $B_c$. 

The total LO $B_c$ and $B_c^*$ cross sections are
\begin{eqnarray}
\label{eqn:sigmabc}
{\sigma}_{Bc} = {42.13}_{-4.88-14.77}^{+6.38+38.14}\,\text{fb}\,,\nonumber \\
\label{eqn:sigmabcstar}
{\sigma}_{B_c^*} = {164.17}_{-18.35-58.65}^{+23.49+151.34}\,\text{fb}\,,
\end{eqnarray}
where the former uncertainty are from quark mass only, and the latter one combined with the former one represent the overall uncertainty. In the case of $B_c$, the mass uncertainty only accounts for a minor fraction of the total uncertainty.  

It is not surprising that the number of $B_c$ events detected by HERA is insufficient. The production cross section of bottom mesons is generally two to three orders of magnitude lower than that of charm mesons, and the reconstruction efficiency is also smaller. Although most of $B_c^*$ decay into $B_c$, the branching ratio of $B_c^\pm \to J/\psi\pi^\pm \approx 0.5\%$ \cite{Chang:1992pt} is too small to generate enough events for reconstruction. At future colliders with higher energy or luminosity, sufficient $B_c$ events might be attainable. The $ep$ colliders under planning include EIC and EicC, both of which are high luminosity colliders with polarized beams. Given that the EicC is designed to probe the small $x$ region of PDF, with electron energy of $5\,\text{GeV}$ and proton energy of $26\, \text{GeV}$, the cross section is $\lesssim 0.03\,\text{fb}$, thus no $B_c$ will be observed at EicC. According to EIC yellow report \cite{AbdulKhalek:2021gbh}, the unpolarized LO cross sections are calculated under electron energy $10\,\text{GeV}$, proton energy $275\,\text{GeV}$. The cut-off is $Q^2 > 1\,\text{GeV}^2$, $p_T > 1\,\text{GeV}$, $20\,\text{GeV} < W < 80\,\text{GeV}$, $0.05 < z < 0.9$, and alternative selections could be made. The results are as follows:
\begin{eqnarray}
\label{eqn:eicsigmabc}
{\sigma }_{Bc}^{\text{EIC}}={5.44}_{-1.03-1.91}^{+0.39+6.36}\,\text{fb}\,,\nonumber\\
    \label{eqn:eicsigmabcstar}
    {\sigma }_{B_c^*}^{\text{EIC}}={22.12}_{-4.37-7.8}^{+1.07+25.6}\,\text{fb}\,.
\end{eqnarray}
Taking luminosity $10^{34}\,\text{cm}^{-2}\text{s}^{-1} \approx 315\,\text{fb}^{-1}\text{year}^{-1}$ and that all $B_c^*$ decays into $B_c$, $3922 \sim 19209$ $B_c$ are produced per year. Again, assuming that $B_c$ is reconstructed through $B_c^\pm \to J/\psi\pi^\pm$, with a branching ratio of 0.5$\%$. Then, for electroproduction, $J/\psi$ can be reconstructed through $J/\psi \to l^+l^-\,(l = e,\,\mu)$, and the branching ratio is 12$\%$ \cite{ParticleDataGroup:2024cfk} for $e$ and $\mu$ combined. Finally, doubling the events since there are $B_c^+$ and $B_c^-$, about $5 \sim 28$ events can be detected per year at EIC. With the following upgrades of high-luminosity EIC (HL-EIC) \cite{Accardi:2012qut}, the luminosity can increase by approximately an order of magnitude and the electron beam energy can reach $16\,\text{GeV}$ without affecting the luminosity. In summary, observing $B_c$ at EIC is promising.   

In the framework of NRQCD, the electroproduction cross sections of $D_s$ and $B_s$ at HERA are also calculated. Due to the light $s$ quark component, the LO cross section and its uncertainty are significantly enhanced compared to $B_c$. Since $m_s$ is about $0.5\,\text{GeV}$ only, the variation in quark mass is constrained within the range of $\pm30\,\text{MeV}$. However, this sensitivity to quark masses may serve as a window for experimental measurements. For example, in the case of $B_s$, although the total mass is 5.4 GeV, a quark mass variation of $0.1\,\text{GeV}$ is sufficient to induce a significant change greater than 200$\%$ in the cross section. 

\begin{figure}[t!]
    \includegraphics[width=0.75\textwidth]{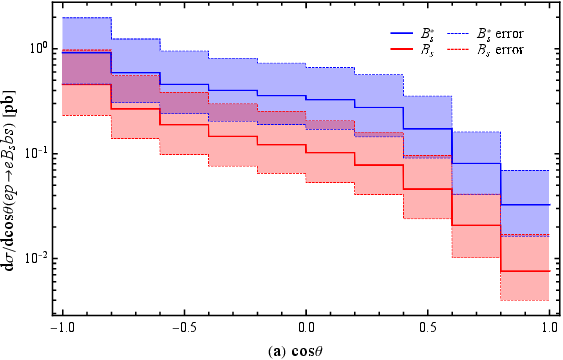}
    \includegraphics[width=0.75\textwidth]{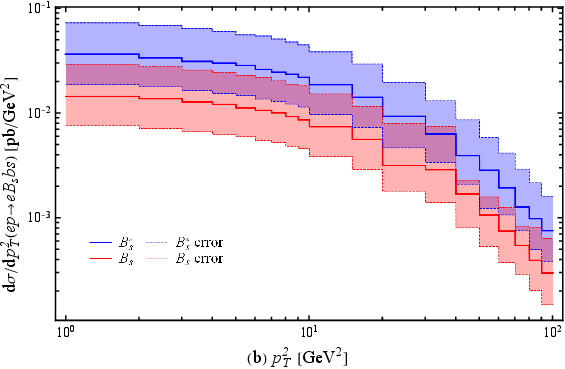}
    \caption{The differential cross section in (a) $\cos\theta$ and (b) $p_T^2$ distributions of the electroproduction process $e+g\to e+B_s+\overline{s}+b$ at LO. Blue lines represent the vector $B_s^*$, and red lines represent the pseudoscalar $B_s^{0}$. The upper bound corresponds to $m_s=0.47\,\text{GeV}$, $m_b=4.93\,\text{GeV}$, and $\mu=\frac{1}{2}m_T$. The lower bound corresponds to $m_s=0.53\,\text{GeV}$, $m_b=4.87\,\text{GeV}$, and $\mu=2m_T$.}
\label{fig:bs}
\end{figure}

\begin{figure}[t!]
    \includegraphics[width=0.75\textwidth]{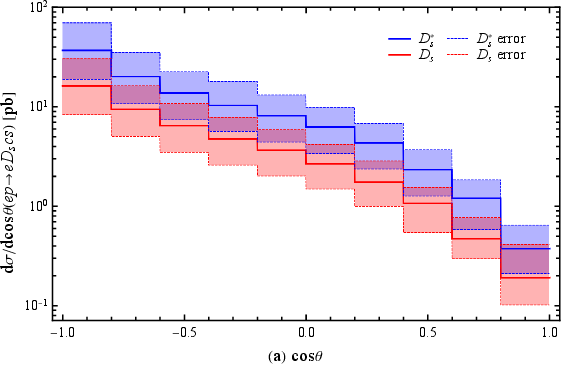}
    \includegraphics[width=0.75\textwidth]{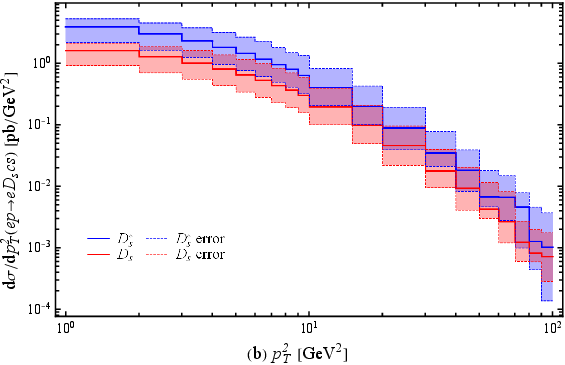}
\caption{The differential cross section in (a) $\cos\theta$ and (b) $p_T^2$ distributions of the electroproduction process $e+g\to e+D_s+\overline{c}+s$ at LO. Blue lines represent the vector $D_s^{*+}$, and red lines represent the pseudoscalar $D_s^+$. The upper bound corresponds to $m_s=0.47\,\text{GeV}$, ${m}{c}=1.53\,\text{GeV}$, and $\mu=\frac{1}{2}m_T$. The lower bound corresponds to $m_s=0.53\,\text{GeV}$, $m_c=1.47\,\text{GeV}$, and $\mu=2m_T$.}
\label{fig:cs}
\end{figure}

The total LO cross sections at HERA are
\begin{eqnarray}
\label{eqn:sigmabs}
{\sigma }_{Bs}={0.29}_{-0.05-0.09}^{+0.06+0.25}\,\text{pb}\,,\nonumber\\
    \label{eqn:sigmabsstar}
    {\sigma }_{B_s^*}={0.73}_{-0.12-0.23}^{+0.15+0.63}\,\text{pb}\,,\nonumber\\
        \label{eqn:sigmacs}
{\sigma }_{Ds}={9.35}_{-1.33-3,03}^{+1.71+5.13}\,\text{pb}\,,\nonumber\\
    \label{eqn:sigmacsstar}
    {\sigma }_{D_s^*}={20.78}_{-2.98-2.8}^{+3.80+11.74}\,\text{pb}\,.
\end{eqnarray}
And the differential cross section distributions are shown in Fig. \ref{fig:bs} and Fig. \ref{fig:cs}. Their trends are analogous to $B_c$ yet with broader error bands. In subsequent studies, the precision can be enhanced by incorporating NLO calculations and more thorough analyses.

Using an integrated luminosity of $\mathcal{L} = 315\,\text{pb}^{-1}$, $48 \sim 189$ $B_s^{0}$, $119 \sim 476$ $B_s^*$, $1575 \sim 5100$ $D_s^+$, $3465 \sim 11440$ $D_s^{*+}$ are produced at HERA. Considering the high reconstruction efficiency, these events are sufficient for analysis of the main decay channels. Nevertheless, for the investigation of rare decay channels including Cabibbo-suppressed modes, more events must be acquired at future colliders.

\section{Conclusion and outlook}

In this paper, the electroproduction of $B_c$, $B_s$, and $D_s$ at $ep$ colliders at LO are investigated within the framework of NRQCD. The results indicate that $B_c$ is observable at EIC. While $B_s$ and $D_s$ events are abundant, their theoretical LO cross sections derived from NRQCD are more dependent on parameters compared to mesons with two heavy flavors. This sensitivity can also present an opportunity in experiment. The differential cross sections in $\cos\theta$ and $p_T^2$ are also presented, and the production is dominant for $p_T^2 \lesssim 30\,\text{GeV}^2$ and $\cos\theta \lesssim -0.4$. The LO uncertainty limits the accuracy of theoretical predictions, which can be reduced in future NLO calculations. At future high-luminosity $ep$ or $ee$ colliders, it is hopeful to accumulate sufficient and clean data for detailed analyses including polarization distributions and rare decay channels, which are crucial for testing NRQCD and exploring new physics.

\section*{Acknowledgements} 

This work was supported in part by the National Natural Science Foundation of China(NSFC) under the Grants 12235008 and 12475087.

\end{document}